\documentstyle[psfig]{caosp}
\def\LB{$\lambda$\,Bootis }
\begin{document}
\pubyear{1993}
\volume{23}
\firstpage{1}
\hauthor{E. Paunzen}
\title{The \LB stars}
\author{E. Paunzen} 
\institute{Institut f\"ur Astronomie der Universit\"at Wien, T\"urkenschanzstr.
17, \\
 A-1180 Wien (paunzen@galileo.ast.univie.ac.at)}
\maketitle
\begin{abstract}
In this article the current knowledge of the group
of \LB stars is reviewed. These metal poor objects are quite outstanding
compared to other chemically peculiar stars of the upper main sequence. Up to
now no theory has been developed which is able to explain the majority
of observational results. This article is mainly focused on the work which
needs to be done in the future in order to clarify the \LB phenomenon.
\end{abstract}

\section{Introduction}

\begin{figure}[hbt]
\centerline{
\psfig{figure=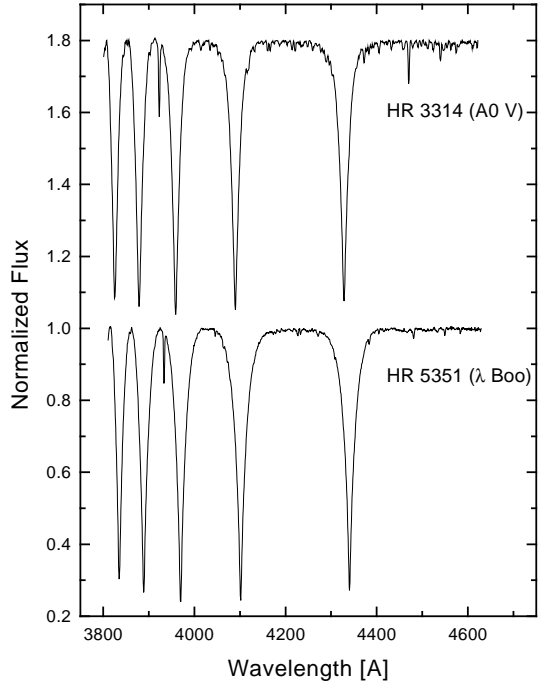,height=11.5cm}}
\caption{Intermediate resolution spectrum of \LB and of a corresponding MK
standard star.}
\label{spectrum}
\end{figure}

This group of metal poor stars was introduced with the identification of an
`abnormal' spectrum for \LB (Figure \ref{spectrum})
in the classification survey of Morgan et al. (1943). Meanwhile,
more stars with the same peculiarities were discovered, 
in particular when the available spectral region was extended 
to the red
and beyond the optical region towards the IR and to the UV.

This flood of additional information is helpful on the one hand in providing
more
physical evidence needed to understand the nature and evolution of this group
of stars,
on the other hand it resulted also in obscuring the group of \LB stars by
objects which definitely have not much in common with the
prototype. Confusion peaked in the eighties, when the group of \LB stars
degenerated to a sort of trash can for stars which could not be classified
otherwise.
This development is described in reviews on \LB stars by Gerbaldi \&
Faraggiana (1993) and Gray (1997).

This was also the reason for a general discussion about the definition
of \LB stars at this workshop. In the following I will give my personal
point of view summarizing the observational facts and proposed theories.

\section{Working definition}

It is evident that a homogeneous and statistically sufficiently large
catalogue of $\lambda$ Bootis stars is required before applying any
statistically sound analysis.
Therefore, the compilation
of such a catalogue (Paunzen \& Gray 1997; Paunzen et al. 1997)
 was the first
goal and a $\lambda$ Bootis classification criterion
had to be extracted from what appears to be a consensus:
\begin{center}
$\lambda$ Bootis stars are Pop.\,I, A- to F-type, metal poor stars, \\
with solar abundances of C, N, O and S.
\end{center}

The spectral types are easily translated into $T_{\rm eff}$ and
log\,$g$ values
which can be estimated fairly accurately with photometric indices or on
spectroscopic grounds. No restriction is chosen for the luminosity
class, because
it remains to be determined in a survey, to what extent the
$\lambda$ Bootis phenomenon is limited to the
main-sequence or includes also evolved stars.
 
Pop.\,I criterion is more problematic.
Evidence have been compiled by Michaud \& Charland
(1986) for this property, however, for a rather small sample. The arguments
were low radial velocities and $U$, $V$ and $W$ velocities which are typical
of Pop.\,I.
We were able to confirm that $\lambda$ Bootis stars have a similar
$v$\,sin\,$i$ distribution (Paunzen
et al. 1997) as is typical of luminosity class {\sc IV} and {\sc V} stars in
our solar
neighbourhood. Periods of
pulsating $\lambda$ Bootis stars are similar to (Pop.\,I) $\delta$\,Scuti
stars (Paunzen
et al.
1998), which indicates that the spectral peculiarities are restricted only to
the stellar
atmosphere and not to the stellar interior. Some authors prefer to substitute
the `Pop.\,I' criterion by
`dwarf', but
that might introduce bias when testing $\lambda$ Bootis theories by
excluding evolved stars
systematically. The same argument applies to `core hydrogen burning stars'.
I have chosen to retain the Pop.\,I criterion which is also
used in the astronomy and astrophysics reference handbook: Landolt-B\"ornstein
(Seitter \& D\"urbeck 1982). Basically, `Pop.\,I' means in the context of the
definition
of $\lambda$ Bootis stars that the observed low metallicity is only a
surface phenomenon of otherwise solar abundant stars.

Several additional peculiarities were found (reviewed by Gerbaldi \&
Faraggiana 1993) for subsets of $\lambda$ Bootis stars which, however,
should not be considered as primary classification indicators, as they are
either
a consequence of the chosen definition, or detectable only for
extreme cases, or are incorrect. An example for the first is
the 160\,nm spectral feature which
is caused by a quasimolecular absorption leading to a satellite in the
Ly$\alpha$ profile
due to perturbation by neutral Hydrogen. Detectability in $\lambda$ Bootis
stars is
possible due to reduced line blending caused by the low metallicity
(Holweger et al.
1994). For the second type of criterion one can refer to an IR excess
observed above
a 2$\sigma$ level with IRAS (King 1994) for only 2 out of the 20
$\lambda$ Bootis stars in the membership lists. An example for the last
type of criterion are the `very large $v$\,sin\,$i$ values' attributed
to $\lambda$ Bootis stars which are
not corroborated by investigations of Abt \& Morrell (1995) and
Paunzen et al. (1997).

Spectral features typical for circumstellar shells have been identified
in five out of eleven
observed $\lambda$ Bootis stars by Holweger \& Rentzsch-Holm (1995).
In at least one case (HD~111786)
evidence is controversial and may be actually caused by an SB2 system
(Faraggiana et al. 1997).

\begin{figure}[hbt]
\centerline{
\psfig{figure=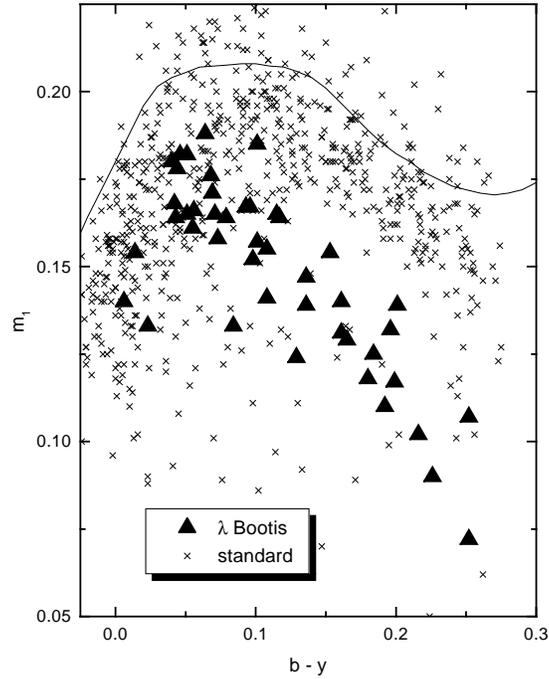,height=11.5cm}}
\caption{{\em m$_{1}$} versus {\em b$-$y}. Crosses are normal stars from Gray 
\& Garrison (1987, 1989a,b), filled triangles are \LB stars from 
Paunzen et al. (1997) and Paunzen \& Gray (1997). The standard line is taken 
from  Philip \& Egret (1980)}
\label{m1}
\end{figure}

\section{Photometric properties}

\begin{figure}[hbt]
\centerline{
\psfig{figure=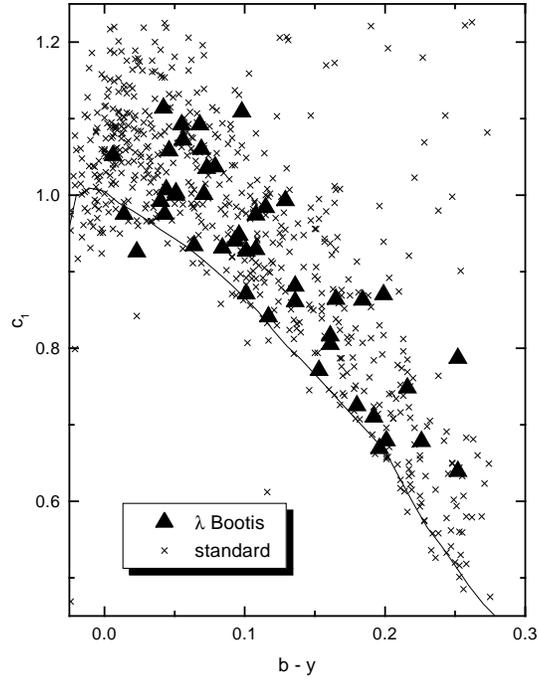,height=11.5cm}}
\caption{{\em c$_{1}$} versus {\em b$-$y}. Symbols are the same as in Fig. 2}
\label{c1}
\end{figure}

Narrow band photometry has often been used to distinguish chemically
peculiar from normal stars.
The Str\"omgren (Fig.\,\ref{m1}, \ref{c1}) as well as the Geneva
photometric systems provide estimates of
temperature, surface gravity and chemical composition of stars.
However, these calibrations were derived for {\it normal}
stars with {\it solar abundances}.
Nevertheless, the conclusions from the Figures mentioned are:
\begin{itemize}
        \item{\it Metallicity:}
                All members have a low metallicity which decreases
                with temperature. A photometric parameter space which
                includes all candidate stars is well determined.
        \item{\it Temperature:}
                The temperature ranges from early A\,V to early F\,V stars.
        \item{\it Surface gravity:}
                \LB stars cannot be distinguished from normal dwarf stars.
        \item{\it Confusion with non-\LB stars:} As is also obvious from
                the Figures,
                other stars populate the same parameter space.
                An unambiguous detection of \LB stars with standard
                photometry alone is
                impossible, although to some extend a discrimination
                with metallicity sensitive indices is useful (e.g. the
                $\Delta$a index was found to be negative for all \LB
                stars by Maitzen \& Pavlovski 1989a,b). But such 
                attempts have led in the past to many spurious
                entries in various catalogues and to confusion in the subject.

\end{itemize}

In parallel to the photometric observations needed for the colour indices,
the photometric stability of the catalogue stars was also 
investigated. 
Altogether, 52 programme stars have been used to analyze the pulsation
properties of \LB stars (Paunzen et al. 1998). 
Using observational ($M_{\rm V}$ vs. $b-y$) and calibrated
($\log(L/L_{\sun})$ vs. $\log T_{\rm eff}$) diagrams, a ratio of
variable to nonvariable members of at least 50\,\% was derived inside the 
classical instability strip. 

The location of all known pulsating programme stars in a
$\log (\rho/\rho_{\sun})$ vs. $\log P$ diagram is
consistent with those of (solar abundant) $\delta$\,Scuti stars which
supports the hypothesis that the \LB abundance pattern (see the article by
U. Heiter in this proceedings)
is restricted to the stellar surface. The possibly
excited modes range from the fundamental mode up to high overtones. Otherwise
no outstanding behaviour compared to $\delta$\,Scuti stars seems to be
present.

These results encourage us to apply the tools of asteroseismology in order
to secure information about the evolutionary status and the {\it overall}
chemical composition of \LB stars.

\section{Other observational results}

In this section I would like to give a summary of observational
results found for \LB stars. Most of these properties are {\it not} shared
by all members. Please note that the number of members is still small
(less than 50), making the conclusions preliminary.
\begin{itemize}
\item{\it Ultraviolet region:} Strong absorption features for some members
(Faraggiana et al. 1990) at 160 and 304\,nm were found. These features are not
unique for this group. IUE low resolution spectra help to distinguish
\LB from FHB stars (Solano \& Paunzen 1998).
\item{\it Optical domain:} No magnetic field was found so far (Bohlender \&
Landstreet 1990). Indications of a gas shell for five out of eleven members
were detected. Pulsation is a common fact for \LB stars inside the classical
instability strip. Three SB systems with \LB type components are known.
\item{\it Infrared region:} There are only two \LB stars with a prominent
IR excess known. Recent ISO results indicate more candidates.
\end{itemize}
For more details refer
to the reviews by Gerbaldi \& Faraggiana (1993) and Gray (1997).

\section{Theories about the \LB phenomenon}

Two competing theories were developed in order to understand the
origin of the $\lambda$ Bootis phenomenon.

The original diffusion/mass loss theory advanced by Michaud \& Charland
(1986) produced $\lambda$ Bootis stars only at the {\it end} of their
main sequence lifetimes. They showed that a mass loss rate of about
10$^{-13}$
solar masses per year together with diffusion in the stellar atmosphere
can actually lead to the observed underabundances of elements after
10$^{9}$
years. According to their theory, $\lambda$ Bootis stars have to be old and
evolved stars. 

Alternatively, Venn \& Lambert (1990) suggested that the $\lambda$
Bootis phenomenon is associated with the accretion of metal-depleted gas from
the interstellar medium. Charbonneau (1991) has shown that this
theory can also produce $\lambda$ Bootis stars, but contrary to the theory of
Michaud \& Charland, these stars would be
{\it unevolved} $\lambda$ Bootis stars
at the ZAMS.
It is of considerable significance that Gray \& Corbally (1993)
discovered the first clear example of a $\lambda$ Bootis star on the ZAMS
of the Orion OB1 association. Recently, further members of this group
were detected in young open clusters such as Blanco~1 and NGC~2264
(Paunzen \& Gray 1997).
These discoveries strengthen the hypothesis that
the $\lambda$ Bootis phenomenon is related to an evolutionary phase
associated
with the arrival of a star on the ZAMS, and in particular this seems to rule
out the diffusion/mass loss theory. This is in accordance with the
suggestion by
King
\& Patten (1992) that $\lambda$ Bootis stars may be
connected to $\beta$ Pictoris and/or Vega-type stars. On the other hand,
the range of stellar ages in which the
$\lambda$ Bootis phenomenon occurs is {\it not} well established.

Very recently, Andrievsky (1997) suggested that \LB stars might be `products'
of W UMa binary systems. According to this theory, \LB stars are remnants
of close (interacting) binary systems resulting in {\it evolved} stars.

With the new Hipparcos data it was possible to establish that at least six
stars are {\it very close to the Zero Age Main Sequence}. Details
can be found in the articles of E. Paunzen and A.E. G\'omez in these
proceedings.

\section{What needs to be done?}

There are some points which are important in order to solve the still open
questions about the \LB phenomenon. In the following I will review some of
them:
\begin{itemize}
\item{\it Homogenity of the group:} Further members have to be established
(on the grounds of as many criteria as possible) in order to improve any
statistics. This has to be done as an iterative process, keeping the
criteria as ``free parameters''. 
\item{\it Common properties:} After one has established a homogenous
group of stars as large as possible, their common properties have to be
studied. An analysis of some parameters (e.g.: $v$\,sin\,$i$ distribution,
behaviour in the infrared, abundance
pattern) might help to clarify the physical processes responsible for this
phenomenon.
\item{\it Evolutionary state:} In order to test any proposed theory it is
essential to know the evolutionary state of this group. Finding \LB stars
in open clusters seems the best way to do so. Calibrations of ages using
(theoretical) models always suffer from approximations (e.g. rotation,
overshooting).
\item{\it Theories:} Using these new boundary conditions, theories might be
improved taking all observational results into account.
\end{itemize}
This workshop has demonstrated an increasing interest in this group. I hope
that many members of our working group are able to contribute in this field.


\acknowledgements
This research was carried out within the working group
{\it Astero-} {\it seismology-AMS } with
funding from the Fonds zur F\"orderung der wissenschaft\-lichen
Forschung (project {\em S7303-AST}). I would like to thank S.M.~Andrievsky,
D.A. Bohlender, B.~Duffee,
R.~Faraggiana, M.~Gerbaldi, R.O.~Gray, G.~Handler, I.Kh.~Iliev,
V.~Malanushenko, D.E.
Mkrtichian, P.~North, I.~Rentzsch-Holm, E.~Solano, A.~Torres-Dodgen
and the Vienna Working
Group for their help during the last years.
Use was made of the Simbad database,
operated at CDS, Strasbourg, France.


\end{document}